  \documentclass[10pt,oneside,a4paper]{article}

  \usepackage[
  backend=biber,
  sorting=nyt
  ]{biblatex}
  \usepackage[pdftex]{graphicx}
  \graphicspath{{../pdf/}{../res/}}
  \DeclareGraphicsExtensions{.pdf,.jpeg,.png}
  \usepackage{amsmath}
  \usepackage{amssymb}
  \usepackage{array}

  \usepackage{pgfplots}
  \usepackage{pgfplotstable}
  \usepgfplotslibrary{fillbetween} 
  \usepackage{tikz}
  \usepackage{caption}
  \usepackage{subcaption}
  
\usepackage{amsmath}
\usepackage{amssymb}

\usepackage{algorithm}

\usepackage{caption}

  \usepackage{algorithmicx}
  \usepackage{algpseudocode}

  \pgfplotsset{width=7cm,compat=1.3}
  \usepackage{csvsimple}
  \usepackage{multirow}
  \usepackage{dcolumn}

  \usepackage{booktabs, makecell}
  \setcellgapes{3pt}
  \usepackage{siunitx} 

  \usepackage{url}

  \usepackage{authblk}
  \usepackage[title]{appendix}%
  \usepackage{comment}

  \usepackage{amsthm} 

  \usepackage{titlesec} 
  \titleformat{\section}
  {\normalfont\Large\bfseries}{\thesection}{1em}{}

\begin{document}

  \title{The Impact of Sequential versus Parallel Clearing Mechanisms in Agent-Based Simulations of Artificial Limit Order Book Exchanges}
  
  \author[1]{Matej Steinbacher\thanks{matej.steinbacher@gmail.com}}
  \author[2]{Mitja Steinbacher\thanks{mitja.steinbacher@kat-inst.si}}
  \author[3]{Matjaž Steinbacher\thanks{corresponding author}}
  
  \affil[1]{%
  Independent Researcher}
  \affil[2]{%
  Faculty of Law and Business Studies, Catholic Institute, Ljubljana, Slovenia}
  \affil[3]{%
  Fund for Financing the Decommissioning of the Krško Nuclear Power Plant and Disposal of Radioactive Waste, Krško, Slovenia}

  \maketitle

  \begin{abstract}
    This study examines the impact of different computing implementations of clearing mechanisms on multi-asset price dynamics within an artificial stock market framework. We show that sequential processing of order books introduces a systematic and significant bias by affecting the allocation of traders' capital within a single time step. This occurs because applying budget constraints sequentially grants assets processed earlier preferential access to funds, distorting individual asset demand and consequently their price trajectories. The findings highlight that while the overall price level is primarily driven by macro factors like the money-to-stock ratio, the market's microstructural clearing mechanism plays a critical role in the allocation of value among individual assets. This underscores the necessity for careful consideration and validation of clearing mechanisms in artificial markets to accurately model complex financial behaviors.
  \end{abstract}


  \section{Introduction}

  Computational simulation has become an indispensable tool within economics and finance, providing a critical methodology for investigating complex, dynamic systems populated by interacting agents \cite{farmer2009economy}, \cite{lebaron2006agent}, \cite{samanidou2007agent}, \cite{tesfatsion2006agent}, \cite{steinbacher2021advances}, \cite{axtell2025agent}. Its value lies in enabling the analysis of phenomena—ranging from agent-based models of financial markets and macroeconomic dynamics to the estimation of complex econometric structures—that are often intractable through analytical methods or direct empirical observation alone. However, ensuring the validity and reliability of simulation-based research hinges critically on the computational model's ability to faithfully represent the theoretical system under study. A fundamental challenge emerges in mapping the inherent simultaneity of many economic and financial processes onto computational frameworks that default to sequential execution.

  While the application of parallel computing in this domain is commonly driven by the pursuit of increased computational speed and the capacity to handle ever-larger models – a focus primarily on efficiency – there exists an equally, if not more, critical dimension concerning the impact of execution strategy on the correctness and structural validity of the simulation results, see, for instance, \cite{axtell2025agent}. Economic and financial interactions, such as simultaneous trading decisions across multiple assets within a defined period, the concurrent assimilation of information by diverse agents, or the coordinated allocation of finite resources under constraints, are conceptually designed as contemporaneous events. Direct translation of these processes into strictly sequential algorithms can introduce computational artifacts that fundamentally distort the simulated dynamics.

  Consider the case of agent-based simulations involving multiple traders interacting across several asset markets, particularly under binding resource constraints like limited trading budgets. A common sequential implementation might process these markets or assets one after another. As illustrated by simulations of this nature, such sequential processing can impose artificial order dependencies that lead to outcomes qualitatively distinct from those where interactions are handled concurrently. For example, sequential processing often results in the arbitrary ordering of asset handling dictating their relative performance and price trajectories, manifesting as a clear separation and ranking of asset prices that is dependent on the processing sequence. This stands in stark contrast to outcomes observed under parallel execution, where the same assets might exhibit much more intertwined dynamics and converge towards similar price levels, reflecting a more accurate representation of agents simultaneously considering multiple investment opportunities within a global budget constraint.

  We argue argue that, compellingly demonstrated by the divergent outcomes observed in sequential versus parallel simulations of multi-asset trading systems, parallel programming is frequently not merely an optimization technique but a methodological prerequisite for achieving rigorous and unbiased simulation results in economics and finance. By enabling the concurrent execution of conceptually simultaneous processes, parallelization can effectively eliminate the order-dependent biases introduced by sequential processing. Our primary focus is thus not on quantifying computational speedup, but on demonstrating how the failure to adopt appropriate parallelization can introduce fundamental incorrectness into simulation results, leading to misleading insights and potentially flawed theoretical or policy conclusions that are artifacts of the computational method rather than genuine properties of the economic system being modeled. We contend that for a significant class of simulation-based research characterized by simultaneous interactions and constraints, parallel computing is essential for ensuring the structural validity and fidelity of the simulation outcomes.

  The study proceeds as follows. Section \ref{sec:model} builds a model that is a base for traders' decision-making. Section \ref{sec:proof_clearing} shows mathematically that clearing mode affects results. A pseudo-code of the computer algorithm is shown in Section \ref{sec:alg}, while simulations are processed and analyzed in Section \ref{sec:simulations}. Last section concludes.

  \section{The Model}
  \label{sec:model}

  The model is a multi-asset version of the \cite{steinbacher2025lob, steinbacher2025bimodal}.\footnote{For a general discussion on limit order book models, see, for instance, \cite{parlour2008limit}, \cite{gould2013limit}, \cite{abergel2016limit}.}
  Let the discrete-time, discrete-state environment be populated with $i \in \{1, 2, ..., N\}$ interacting agents, who trade multiple assets $S = S_1, ..., S_J$ over a time horizon $t \in \{0, 1, ..., T\}$ and whose wealth $W_t^i = M_t^i + \sum_{j=1}^{J}\left( S_{j,t}^i P_{j,t} \right), \quad i \in \{1, 2, ..., N\}$ in time $t$ is composed of two components: monetary holdings $M_t^i \in \mathbb{R}{\ge 0}$ and holdings of assets $S_{j,t}^i$ with $P_{j,t} \in \mathbb{R}^+$ is the market price of the asset $j$ at time $t$. Since short selling through negative money holdings is not allowed and $S_{j, t}^i \in \mathbb{Z}^+$ which means that only non-negative integer position in the asset is allowed.

  Agents in the model trade the stock by maximizing their expected wealth of the form 
  \begin{equation}
    \max_{\{\pi_t^i\}_{t=0}^{T-1}} \mathbb{E}_0\left[\sum_{t=1}^{T} \beta^{t-1} u(W_t^i)\right],
    \label{eq:objective}
  \end{equation}
  where $\pi_t^i$ represents the trading strategy of agent $i$ at time $t$; $\beta \in (0, 1)$ is the discount factor, reflecting the time preference of the agent; $u: \mathbb{R} \rightarrow \mathbb{R}$ is a utility function, CRRA in our case, which is strictly increasing, concave, and twice continuously differentiable of the form: $u(W) = \frac{W^{1-\gamma}}{1-\gamma}$ for $\gamma \neq 1$ and $u(W) = \ln(W)$ for $\gamma = 1$, where $\gamma > 0$ is the coefficient of relative risk aversion; $\mathbb{E}_0[\cdot]$ denotes the expectation operator conditional on the information available at time $t=0$ and $t=1,2,...,T$ is time domain.

  It was shown in \cite{steinbacher2025lob} that solving the utility function for a myopic, random trader who does not incorporate risk consideration into the trading decision  his trading decision-making to 
    
    \begin{align}
      \label{eq:agent_choice}
      \pi_{t} &= 
      \begin{cases} 
            B_t & P_t^{\ast} < E[P_{t+1}], \\
            A_t & P_t^{\ast} > E[P_{t+1}], \\
            \text{Nothing} & \text{else}
      \end{cases}
  \end{align}
    
  This means that the trader submits an ask order if his private expectation of the price-change is a drop and a bid order if the price is expected to rise. The price expectations are created as
  \begin{equation}
    \label{eq:expected_price}
    E_t^i[P_{t+1} | P_t] = P_t \cdot (1 + \Delta_i),
  \end{equation}
  where $\Delta_i \sim U(-\sigma,\sigma) $ is independently and identically distributed (i.i.d.) from a uniform distribution $U(-\sigma, \sigma)$.

  All orders of an asset $j$ are submitted to the order book $O_j$ that works as a collection of all bids $B_{j, t} = \{ (P^i_b, q^i_b) : P_b \in \mathbb{R}^+, q_b \in \mathbb{N}, i \in \{1, 2, ..., N\} \}$ and asks $A_t = \{ (P^i_a, q^i_a) : P_a \in \mathbb{R}^+, q_a \in \mathbb{N}, i \in \{1, 2, ..., N\} \}$ of the asset $j$.

  $O_j$ operates as a collector of bids and asks and a matchmaker that matches and settles the best bid and ask orders in a form of a trade. The best bid price is the highest bid price available in the market at time $t$ or $P_t^b = \sup \{ P_b : (P^i_b, q^i_b) \in B_t \}$, while the best ask price is the lowest ask price available at time $t$ or $P_t^a = \inf \{ P_a : (P^i_a, q^i_a) \in A_t \}$, where $\sup$ denotes the supremum or least upper bound and $\inf$ denotes the infimum that is greatest lower bound.
  
  A trade is settled at the mid-price $P_{(A,B)} = (P_A + P_B) / 2$. For the trader $i$ this means that the change in cash $\Delta M_{i,t}$ and change in stock holdings $\Delta S_{j,i,t}$ are promptly modified as

  \begin{align*} 
    M_{i,t}(j) &= M_{i,t} + ( Q_{j,k} \cdot P_{j,k,t}(A,B) - Q_{j,l} \cdot P_{j,k,t}(A,B)) \\ 
    S_{i,t}(j) &= S_{i,t,j} +  Q_{j,B} - Q_{j,A} 
  \end{align*}
  where $B$ indicates the trader as a buyer and $A$ as a seller. 

  Finally, a closing price of each time $t$ is set as the price of the last executed trade at $t$ or $P_t = P_t^{\text{last}}$. If no trade occured at time $t$, the market price is assumed to remain unchanged from the previous period: $P_t = P_{t-1}$.


  \section{Proof that Clearing Mode Affects Results}
  \label{sec:proof_clearing}
  
  This section provides the mathematical formalism underlying the multi-asset trading simulation described, specifically highlighting how sequential processing can introduce biases compared to a parallel execution approach, focusing on the implications of hard budget constraints.
  
  Consider $N$ traders and $J$ assets at a fixed time $t$ and assume:
  
  \begin{enumerate}
    \item All asset mid-prices at the start of the tick are equal to a common value $P>0$:
      \[
        P_1(t)=P_2(t)=\cdots=P_J(t)=P.
      \]
      (This assumption isolates the budget-shrinkage effect; the argument extends mutatis mutandis when prices differ, by comparing the relevant budget-threshold events asset-by-asset.)
    \item Each trader submits at most one unit per asset, so orders have size 1.
    \item The market is deep enough that every submitted buy (sell) finds a matching sell (buy), subject only to the trader's own budget/inventory constraint.
    \item Trader $i$ draws i.i.d.\
      \(\varepsilon_{i,j}\sim\mathrm{U}(-\sigma,\sigma)\) for each asset $j$.
      Trader attempts a buy of one unit on $j$ iff $\varepsilon_{i,j}>0$, else a sell.
    \item Cash budgets $B_i>0$ are identical across parallel and sequential regimes at the start of the tick.
  \end{enumerate}
  
  Define the indicator
  \[
    X_{i,j}
    =
    \begin{cases}
      1, & \text{if trader $i$ successfully buys 1 unit of asset $j$,}\\
      0, & \text{otherwise.}
    \end{cases}
  \]
  (Since buys and sells must match one-for-one, the same argument holds for sell volumes.)
  
  \medskip
  
  \noindent\textbf{1.\ Parallel Clearing.}
  Each trader's available cash for asset $j$ is always $B_i$. Thus
  \[
    X_{i,j}
    = \mathbf{1}\{\varepsilon_{i,j}>0\}\,\mathbf{1}\{B_i \ge P\}.
  \]
  Because $\varepsilon_{i,j}$ is independent of budgets and
  \(\Pr(\varepsilon_{i,j}>0)=\tfrac12,\)
  we have
  \[
    \Pr(X_{i,j}=1)
    = \tfrac12\,\mathbf{1}\{B_i \ge P\},
  \]
  which does not depend on the asset index $j$. Hence for the total buy-volume
  \[
    V_j
    \;=\;
    \sum_{i=1}^N X_{i,j},
  \]
  we get
  \[
    \mathbb{E}[V_j]
    = \sum_{i=1}^N \Pr(X_{i,j}=1)
    = \sum_{i=1}^N \frac12\,\mathbf{1}\{B_i \ge P\}
    \quad\text{(independent of $j$).}
  \]
  Thus
  \[
    \mathbb{E}[V_1]
    = \mathbb{E}[V_2]
    = \cdots
    = \mathbb{E}[V_J].
  \]
  
  \medskip
  
  \noindent\textbf{2.\ Sequential Clearing.}
  Trader $i$'s residual cash before trading asset $j$ is
  \[
    B_{i,j-1}
    \;=\;
    B_i
    \;-\;
    \sum_{k=1}^{j-1} P\,X_{i,k},
    \quad
    j=1,\dots,J,
  \]
  with $B_{i,0}=B_i$. The trader buys an asset $j$ iff
  \[
    \varepsilon_{i,j}>0
    \quad\text{and}\quad
    B_{i,j-1}\ge P.
  \]
  Hence
  \[
    X_{i,j}
    = \mathbf{1}\{\varepsilon_{i,j}>0\}\,\mathbf{1}\{B_{i,j-1}\ge P\}.
  \]
  
  Because $\varepsilon_{i,j}$ is independent of $B_{i,j-1}$,
  \[
    \Pr(X_{i,j}=1 \mid B_{i,j-1})
    = \Pr(\varepsilon_{i,j}>0)\,\mathbf{1}\{B_{i,j-1}\ge P\}
    = \tfrac12\,\mathbf{1}\{B_{i,j-1}\ge P\}.
  \]
  Taking unconditional probability gives
  \[
    p_{i,j}
    := \Pr(X_{i,j}=1)
    = \tfrac12\,\Pr\bigl(B_{i,j-1}\ge P\bigr).
  \]
  
  We now show $p_{i,j}<p_{i,j-1}$ for each $j\ge2$.
  
  \medskip
  
  \noindent\textbf{Lemma.} For $j\ge2$,
  \[
    \Pr\bigl(B_{i,j-1}\ge P\bigr)
    \;<\;
    \Pr\bigl(B_{i,j-2}\ge P\bigr).
  \]
  
  \begin{proof}
  Since $X_{i,j-1}\in\{0,1\}$,
  \[
    B_{i,j-1}
    = B_{i,j-2}
    - P\,X_{i,j-1}
    =
    \begin{cases}
      B_{i,j-2} - P, & \text{if }X_{i,j-1}=1,\\
      B_{i,j-2},     & \text{if }X_{i,j-1}=0.
    \end{cases}
  \]
  Hence
  \[
    \{\,B_{i,j-1}\ge P\}
    =
    \bigl\{X_{i,j-1}=0,\;B_{i,j-2}\ge P\bigr\}
    \;\cup\;
    \bigl\{X_{i,j-1}=1,\;B_{i,j-2}\ge 2P\bigr\}.
  \]
  These two events are disjoint, so by the law of total probability,
  \[
    \Pr(B_{i,j-1}\ge P)
    = \Pr(X_{i,j-1}=0)\,\Pr(B_{i,j-2}\ge P)
      \;+\;
      \Pr(X_{i,j-1}=1)\,\Pr(B_{i,j-2}\ge 2P).
  \]
  But
  \(\Pr(X_{i,j-1}=0)=1-p_{i,j-1}<1\)
  and
  \(\Pr(B_{i,j-2}\ge 2P)\le\Pr(B_{i,j-2}\ge P).\)
  Therefore
  \[
    \Pr(B_{i,j-1}\ge P)
    < \Pr(B_{i,j-2}\ge P)\,\bigl[(1-p_{i,j-1}) + p_{i,j-1}\bigr]
    = \Pr(B_{i,j-2}\ge P),
  \]
  establishing the desired strict inequality.
  \end{proof}
  
  \medskip
  
  It follows that
  \[
    p_{i,j}
    = \tfrac12\,\Pr(B_{i,j-1}\ge P)
    < \tfrac12\,\Pr(B_{i,j-2}\ge P)
    = p_{i,j-1},
    \quad
    j=2,\dots,J.
  \]
  Summing over $i=1,\dots,N$, the total expected volume
  \[
    \mathbb{E}[V_j]
    = \sum_{i=1}^N p_{i,j}
  \]
  satisfies
  \[
    \mathbb{E}[V_1]
    > \mathbb{E}[V_2]
    > \cdots
    > \mathbb{E}[V_J].
  \]
  
  \medskip
  
  \noindent\textbf{3.\ Price Implications.}
  Assume each executed buy of one unit on asset $j$ transacts at mid-price and updates the closing price
  \[
    C_j = P + \Delta_j(V_j),
    \quad\text{where }\Delta_j(\cdot)\text{ is strictly increasing}.
  \]
  Then the strict ordering of $\mathbb{E}[V_j]$ in sequential clearing implies
  \[
    \mathbb{E}[C_1]>\mathbb{E}[C_2]>\cdots>\mathbb{E}[C_J],
  \]
  whereas under parallel clearing $\mathbb{E}[C_1]=\cdots=\mathbb{E}[C_J]$.
  
  \qed
  

  \section{The Algorithm}
  \label{sec:alg}

  The simulation code was writen in C++ and this section describes the computing implementation of the system with particular focus on a process for advancing the simulation by one time step $t \to t + 1$ for both sequential and parallel execution strategies.

  The state of the simulation at time $t$ is defined by the state of each agent $i \in \{1, \dots, N\}$, $W_{i,t}(M,S_j)$, and the state of each asset market $j \in \{1, \dots, J\}$, like order book $\mathcal{O}_{j,t}$, price $P_{j,t}$. The process for advancing the simulation by one time step $t \to t+1$ is described algorithmically below.

  \subsection{Order Submission Phase}
  During the first phase traders generate and submit orders to the accompanying order books. The process is shown in the Algorithm \ref{alg:condensed_order_phase}. 

  \begin{algorithm}
    \caption{Condensed Order Phase}
    \label{alg:condensed_order_phase}
    \begin{algorithmic}[1]
        \For{each time step $t$}
            \State $\mathcal{O}_t \leftarrow \emptyset$ \Comment{Orders for step $t$}
            \For{each agent $i \in \{1, \dots, N_T\}$}
              \For{each asset $j \in \{1, \dots, J\}$ }
                \State $o_{j,i,t} \leftarrow \text{Agent}_i.\text{generateOrder}(M_{i,t}, S_{j,t}, E_t^i(P_{t+1, j}^i | P_{t,j}))$
                \Comment{Agent $i$ generates order for asset $j$}
                \State $\mathcal{O}_{j,i,t} \leftarrow \text{Agent}_i.\text{submitOrder}(o_{j,i,t})$
                \Comment{Submit order to relevant $\mathcal{O}_j$}
              \EndFor
            \EndFor
        \EndFor
    \end{algorithmic}
  \end{algorithm}

  At the beginning of time step $t = 1, \dots, T$, each trader $i$ observes his current wealth $W_{i,t}$ and the current market state, like prices $P_{j,t}$, generates orders for assets $j \in J$ in a sequential order according to Equations \ref{eq:agent_choice} and \ref{eq:expected_price} and submits them to the accompanying order book $\mathcal{O}_{j,i,t}$. The order of operations is sequential at the level of iteration across traders and assets. Given Equations \ref{eq:agent_choice} and \ref{eq:expected_price} none of the operation orders from the Algorithm \ref{alg:condensed_order_phase} affects the outcome since they are generated independently from one another and with whole trader's wealth available at the time.
  
  This phase is shared by both sequential and parallel clearing modes as discussed in the Section \ref{sec:clearing_order_phase}.

  \subsection{Clearing Order Phase}
  \label{sec:clearing_order_phase}

  \subsubsection{Sequential Clearing}
  Processing of order books $\mathcal{O}_j$ in a sequential mode performs order matching and clearing across assets iteratively, one after another in a fixed order that is set by the asset's index $j = 1,2,...,J$ at each time $t$. The pseudo-code of the algorithm is shown in Algorithm \ref{alg:sequential_clearing}.

  \begin{algorithm}[ht]
    \caption{Simulation Loop: Sequential Clearing}
    \label{alg:sequential_clearing}
    \begin{algorithmic}[1]
      \For{$\text{iteration}=1$ to $T$}
        \For{each asset $j \in J$}
          \State $\mathcal{O}_j \gets \mathrm{getOrderBook}(j)$
          \State initialize $k=0$ \Comment{Reset trade counter}
          \While{$\mathrm{existPossibleTrade}(\mathcal{O}_j)$}
            \State $\mathit{trade}\gets\textsf{matchOneUnit}(\mathcal{O}_j)$
            \State $\{0,1\} \gets \mathrm{validateTrade}(\mathit{trade})$
            \State $\mathrm{finalizeTrade}(\mathcal{O}_j,\mathit{trade})$
            \State $k \gets k+1$
            \State $P_{j,k} \gets \mathit{trade}.\text{price}$        \Comment{Update settlement price}
          \EndWhile
          \State $P_{j,t} \leftarrow P_{j,k}$ \Comment{Update closing price}
          \State $\mathcal{O}_{j,t+1} \leftarrow \emptyset$ \Comment{Clear unfilled orders}
        \EndFor
      \EndFor
    \end{algorithmic}
  \end{algorithm}

  The Algorithm \ref{alg:sequential_clearing} describes a sequential approach to market clearing within each simulation time step $t$. For every iteration (time step $t$), the algorithm processes the markets for each asset $j$ in a predefined sequential order that is index-based.

  The method $\mathrm{validateTrade}(\mathit{trade})$ checks if each trader's strategy respects physical and market constraints based on his state $W_{i,t}$ to close the trade. In particular, $Q_{o_k}(i,t,j) \leq S_{i,t,j}$ for the seller and $Q_{o_k}(i,t,j) \cdot P(o_k) \leq M_{i,t}$ for the buyer, with only approved trades getting cleared.

  \subsubsection{Parallel Clearing}
  A pseudo-code of the algorithm for processing of orders in a parallel mode at the "per-asset" level is shown in Algorithms \ref{alg:parallel_clearing} and \ref{alg:parallel_clearing_pb} that outline the main simulation loop.

  \begin{algorithm}[ht]
    \caption{Simulation Loop: Parallel Clearing}
    \label{alg:parallel_clearing}
    \begin{algorithmic}[1]
      \For{$\text{iteration}=1$ to $T$}
        \State initialize thread pool $\mathit{Pool}$ with $J$ workers
        \For{each asset $j$ in $1,\dots,J$}
          \State $\mathit{Futures}[j]\gets \mathit{Pool}.\textsf{submit}(\textsf{processBook}(j))$
        \EndFor
        \State $\mathit{Pool}.\textsf{join}()$
        \For{each asset $j$ in $1,\dots,J$}
          \State $P \gets \mathit{Futures}[i].\textsf{get}()$
          \State \textsf{recordPrice}($j,P$)
        \EndFor
      \EndFor
    \end{algorithmic}
    \end{algorithm}
    
  \begin{algorithm}[ht]
      \caption{processBook($j$): Per-Asset Clearing}
      \label{alg:parallel_clearing_pb}
      \begin{algorithmic}[1]
        \State $\mathcal{O}_j\gets \textsf{getOrderBook}(j)$
        \State initialize $k=0$ \Comment{Reset trade counter}
        \While{$\mathrm{existPossibleTrade}(\mathcal{O}_j)$}
          \State \textbf{lock}($\mathit{managerMutex}$)
            \State $\mathit{trade}\gets\textsf{matchOneUnit}(\mathcal{O}_j)$
            \State $\{0,1\} \gets \mathrm{validateTrade}(\mathit{trade})$
            \State $\mathrm{finalizeTrade}(\mathcal{O}_j,\mathit{trade})$
            \State $k \gets k+1$
            \State $P_{j,k} \gets \mathit{trade}.\text{price}$        \Comment{Update settlement price}
          \State \textbf{unlock}($\mathit{managerMutex}$)
        \EndWhile
        \State $P_{j,t} \leftarrow P_{j,k}$ \Comment{Update closing price}
        \State $\mathcal{O}_{j,t+1} \leftarrow \emptyset$ \Comment{Clear unfilled orders}
      \end{algorithmic}
  \end{algorithm}
  
  The Algorithm \ref{alg:parallel_clearing} manages the parallel execution flow, distributing the task of clearing each asset's market (defined by the Algorithm \ref{alg:parallel_clearing_pb}) across multiple threads. The Algorithm \ref{alg:parallel_clearing_pb} handles the iterative matching and execution within a single order book, performing state updates concurrently with other assets, synchronized by a mutex. This parallel structure allows the effects of trading across different assets to be processed within the same time step without the artificial sequencing bias introduced by the sequential approach, despite using the same updating mechanism as under the Algorithm \ref{alg:sequential_clearing}, while the mutex ensures safe access to shared agent and market states.

  \section{Simulations}
  \label{sec:simulations}
  The general simulation setup is configured as follows. The agent population is set to $N = 10,000$ traders, each with initial wealth endowment $W_0^i(M^i,S_j^i)$ comprising of monetary holdings $M^i = 200$ and stock holdings $S_j^i = 10$ for $j={1,...,5}$. Initial price for all stocks is set to $P_{j,0} = 10$ for all $j = 1,...,J$. Stocks are traded in a unit and cannot be shorted and traders are a subject of a hard-budget constraint, forbidding them to trade on a margin. Money does not earn any interest.

  The trading horizon spans for $t=1,...,5,000$ discrete time periods. In each iteration $t$, a myopic, random (zero intelligent)\footnote{Traders following zero intelligence (ZI) strategy by placing random order submissions have been widely used for the study of LOB, for instance, \cite{gode1993allocative} or \cite{farmer2005predictive}.} trader makes a trading decision based on Equation \ref{eq:agent_choice} where $E(P_{j,t+1} | I_{j,t}, P_{j,t}) = P_{j,t} \cdot (1 + \Delta \sim \mathbb{U}(-0.15,0.15))$ with $\Delta$ a uniform random variable on the interval. 
  
  Once submitted orders can not be modified. The orderbook, as a match-maker, iteratively matches best bids with best asks, that is the highest-priced bid and the lowest-priced ask, until such pairs can be formed and settles them at the mid-price of the bid's and the ask's price. Any unmatched order at the end of the time interval $t$ is cancelled and discarded. Trading does not involve any costs. Settlements are promptly done. A settled price of the last trade of an iteration is recorded as the closed price of the iteration $P_t$ that becomes public knowledge available to all traders.\footnote{The source code of the artificial stock-market environment was written in C++ and was compiled for the Xcode. In the context of an object-oriented framework of the C++, a trader is modeled as an object with a lifetime and a scope that encapsulates both its state (private and public data) and its behavior (actions). In essence, the C++ trader object acts as a software representation of an autonomous agent operating within the simulated stock market, with its own internal characteristics and a defined set of actions it can take to interact with the market and other agents.}

  \subsection{Sequential Clearing}
  
  \begin{figure}[h]
    \centering
    \includegraphics[width=\textwidth]{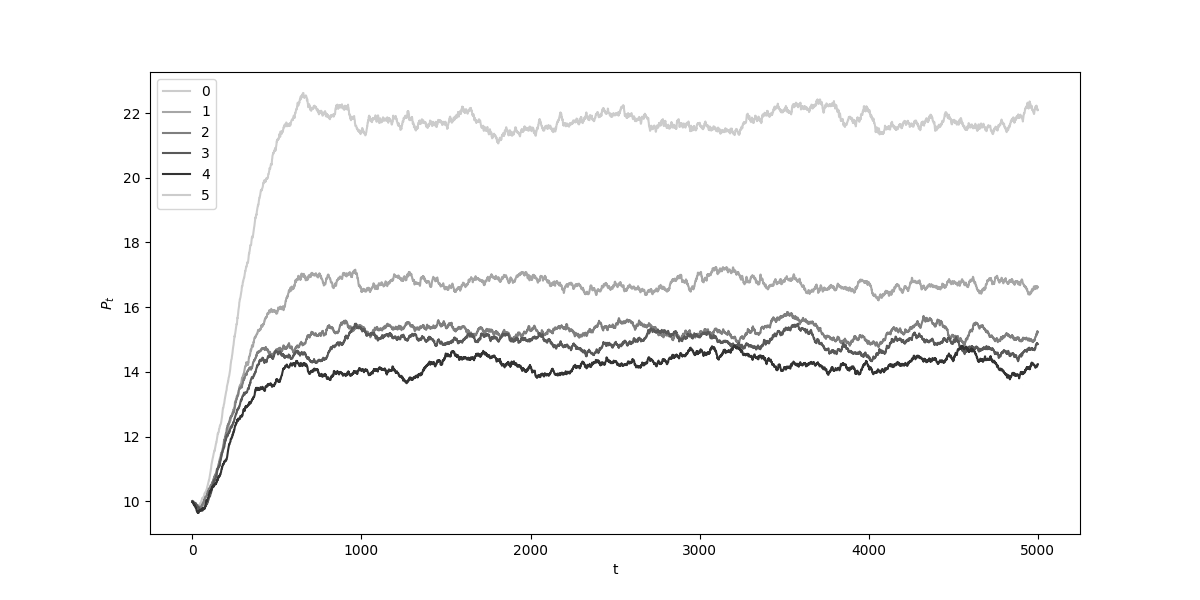}
    \caption{Simulation Results: Sequential Programming}
    \label{fig:simulations_sequential}
  \end{figure}

  Figure \ref{fig:simulations_sequential} depicts price trajectories of assets under the sequential clearing process with the order of processing following the alphabetic order of asset's labels. Asset A consistently reaches the highest price, followed predictably by B, C, D, and E, maintaining a clear hierarchy throughout the simulation period. This outcome aligns directly with the potential bias discussed earlier, where the sequential processing order (e.g., processing trades for A before B, B before C, etc.) can create an systematic advantage for assets appearing earlier in the sequence, particularly when agents face binding constraints like a limited budget that is depleted sequentially. The resulting ordered price levels appear to be an artifact of the simulation's processing structure rather than solely an outcome of the economic interactions themselves.

  The average terminal price $\overline{P_{T}} = 16.61$ with the standard deviation $\sigma_{P_T} = 3.188$. Augmented Dickey-Fuller (ADF) Test (all $\text{p-values} \approx 0.0$ ) indicates that all price trajectories exhibit stationarity which aligns with the visual observation that the prices fluctuate around a relatively constant mean after an initial period. However, the Kruskal-Wallis test (statistic=18184.02, pvalue=0.0) and the Bartlett test (statistic=7520.24, pvalue = 0) indicate that there are statistically significant differences in the central tendency among at least some of the five trajectories in this set, which, again aligns with the visual inspection of the previously provided plot where the trajectories stabilize around different price levels.

  \subsection{Parallel Clearing}
 
  \begin{figure}[h]
    \centering
    \includegraphics[width=\textwidth]{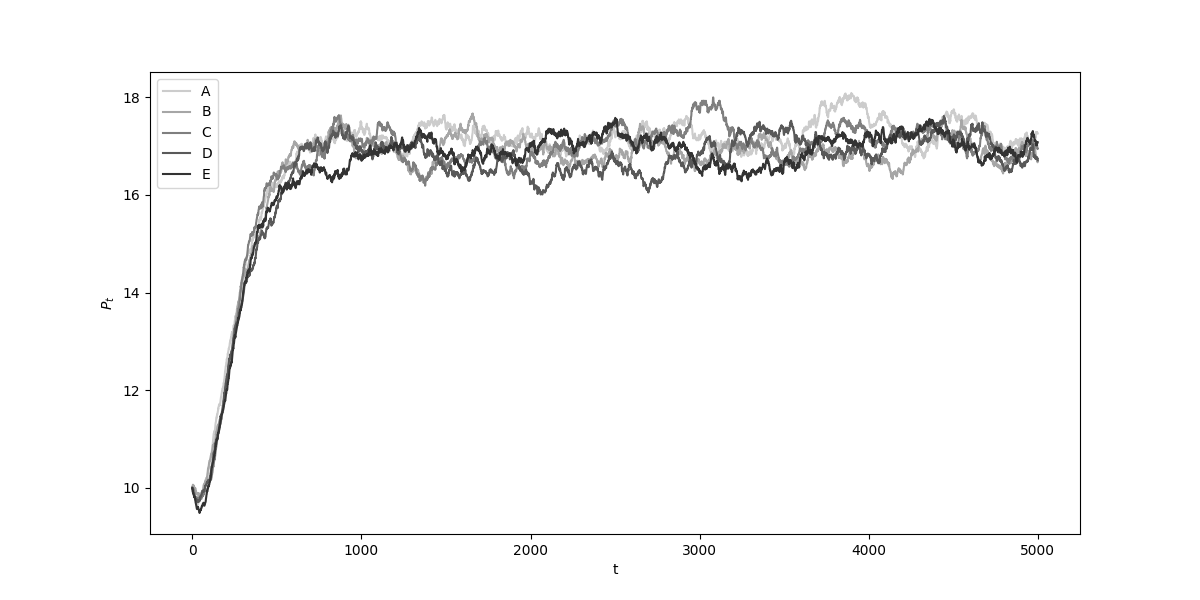}
    \caption{Simulation Results: Parallel Programming}
    \label{fig:simulations_parallel}
  \end{figure}

  In stark contrast, the Figure \ref{fig:simulations_parallel}, depicting simulation results where orders are processed in parallel as they appeared, irrespective of the asset's label, shows the price trajectories for assets A through E remaining much closer together. While there are fluctuations and temporary differences, the prices generally converge to similar levels and exhibit much more intermingled movement. This result is intuitively more consistent with a market where multiple assets are traded by the same agents under global constraints (like total budget) within the same time step, and where information or trading pressure on one asset might quickly influence others without an artificial sequential lag imposed by the simulation algorithm.

  The average terminal price $\overline{P_{T}}=16.94$ which is slightly larger than at the sequential clearing but with a significantly more compresed prices $\sigma_{P_T} = 0.241$. Similarly to the sequential clearing, Augmented Dickey-Fuller (ADF) Test (all $\text{p-values} \approx 0.0$) indicates that all price trajectories exhibit stationarity, while the Kruskal-Wallis test (statistic=2417.71, pvalue=0.0) and the Bartlett test (statistic=17.17, pvalue = 0.00179) indicate the statistically significant differences in the central tendency of the price levels in this set as well (although to a much smaller degree).

  Anyway, a common observation is that the shape of price trajectories, irrespective of the LOB clearing mechanism, demonstrates phases analogous to those observed in single-asset trading: (i) initial drop, (ii) rapid price discovery, (iii) convergence, and (iv) run-specific divergence and stochasticity (\cite{steinbacher2025lob}).


\section{Conclusion}
This study investigated the influence of clearing mechanisms on price trajectories within a multi-asset artificial stock market. We showed, visually and mathematically, how the choice of the clearing mechanism, either a sequential or a parallel, affects the allocation of resources and price discovery.

By processing assets one by one, the sequential method imposes an order-dependent depletion of traders' available capital within a single time step. This artificial constraint grants preferential access to capital for assets processed earlier in the sequence, fundamentally biasing the demand faced by individual assets. Consequently, the sequential mechanism does not merely manage the re-allocation of money to assets according to market forces and initial budgets but it actively distorts this re-allocation based on an arbitrary processing order by granting to assets processed earlier in the sequence a preferential access to the traders' initial capital within that time step compared to assets processed later.

The findings underscore that while the macro-level ratio of money to stock fundamentally influences the general price level of stocks, the microstructural details of the clearing mechanism critically impact how this value is distributed among individual assets. The path dependency and biased allocation inherent in sequential processing highlight that such mechanisms are not neutral with respect to price formation. This study emphasizes the necessity of carefully designing artificial market clearing mechanisms to accurately reflect the intended application of agent constraints and ensure that emergent macro-level properties arise from unbiased micro-level interactions, thereby enhancing the fidelity and reliability of simulation outcomes.

\section*{Disclosure of interest}
The authors declare that they have no conflict of interests relevant to this publication.

\section*{Funding}
No funding was received for this study.

\printbibliography

\end{document}